\pdfoutput=1
\documentclass[11pt]{article}

\usepackage{acl}

\usepackage{times}
\usepackage{latexsym}

\usepackage{amsmath, amssymb, amsthm}

\usepackage[T1]{fontenc}

\usepackage[utf8]{inputenc}

\usepackage{microtype}

\usepackage{enumitem}
\usepackage{subcaption}
\usepackage{cleveref}
\usepackage{graphicx}
\usepackage{multirow}
\usepackage{booktabs}
\usepackage[normalem]{ulem}
\usepackage{algorithm}
\usepackage[noend]{algorithmic}

\newtheorem{definition}{Definition}

\title{A Customized Text Sanitization Mechanism with Differential Privacy}

\author{Huimin Chen$^1$\thanks{\;\;Equal contribution.}, \; Fengran Mo$^{2}$\footnotemark[1], \; Yanhao Wang$^1$, \; Cen Chen$^1$\thanks{\;\;Corresponding author.}, \\ {\bf Jian-Yun Nie$^2$, \; Chengyu Wang$^3$, \; Jamie Cui$^4$} \\
  $^1$East China Normal University \quad $^2$University of Montreal \quad $^3$Alibaba Group \quad $^4$Ant Group \\
  \texttt{saichen@stu.ecnu.edu.cn}, \; \texttt{fengran.mo@umontreal.ca} \\ \texttt{\{yhwang,cenchen\}@dase.ecnu.edu.cn}, \; \texttt{nie@iro.umontreal.ca} \\ \texttt{chywang2013@gmail.com}, \; \texttt{jamie.cui@outlook.com} \\}

\begin{document}

\maketitle

\begin{abstract}
As privacy issues are receiving increasing attention within the Natural Language Processing (NLP) community, numerous methods have been proposed to sanitize texts subject to differential privacy. However, the state-of-the-art text sanitization mechanisms based on metric local differential privacy (MLDP) do not apply to non-metric semantic similarity measures and cannot achieve good trade-offs between privacy and utility. To address the above limitations, we propose a novel \underline{Cus}tomized \underline{Text} (CusText) sanitization mechanism based on the original $\epsilon$-differential privacy (DP) definition, which is compatible with any similarity measure.
Furthermore, CusText assigns each input token a customized output set of tokens to provide more advanced privacy protection at the token level.
Extensive experiments on several benchmark datasets show that CusText achieves a better trade-off between privacy and utility than existing mechanisms.
The code is available at \url{https://github.com/sai4july/CusText}.
\end{abstract}
\section{Introduction}

In many Natural Language Processing (NLP) applications, input texts often contain sensitive information that can infer the identity of specific persons~\cite{jegorova2021survey}, leading to potential privacy leakage that impedes privacy-conscious users from releasing data to service providers~\cite{carlini2019secret, carlini2021extracting, song2020information}.
Moreover, legal restrictions such as CCPA\footnote{\url{https://oag.ca.gov/privacy/ccpa}} and GDPR\footnote{\url{https://data.europa.eu/eli/reg/2016/679/oj}} may further limit the sharing of sensitive textual data.
This makes NLP service providers difficult to collect training data unless the privacy concerns of data owners, including individuals and institutions, are well discoursed.

\begin{figure}[t]
  \centering
  \includegraphics[width=\linewidth]{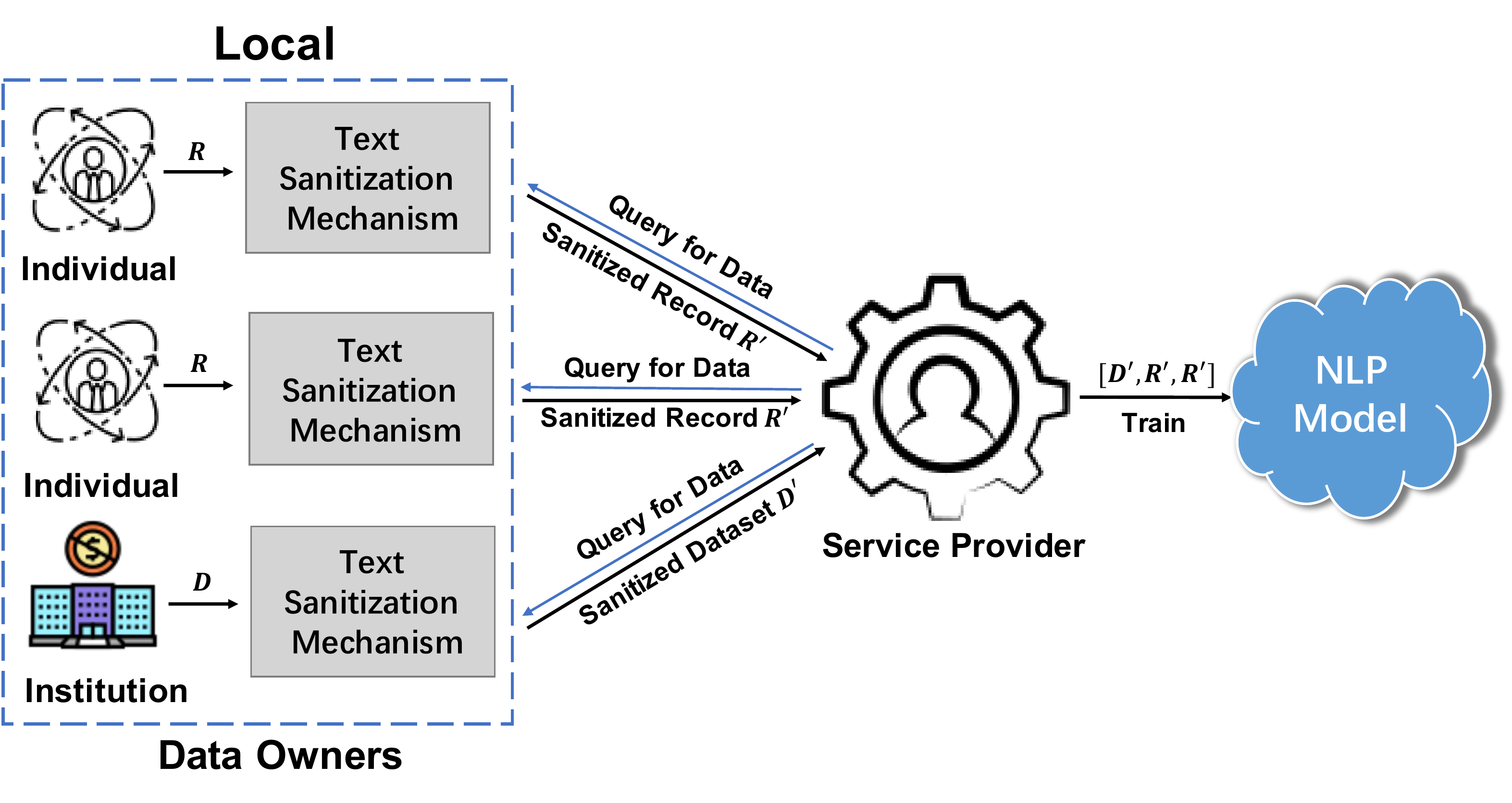}
  \caption{A privacy-preserving NLP workflow.}
  \label{fig:prac}
\end{figure}

To address such privacy issues, great efforts~\cite{lyu2020differentially, anil2021large, dupuy2021efficient, li2021large, mireshghallah2021privacy} have been made to train language models (LMs) with \emph{differential privacy}~\cite{dwork2006calibrating} (DP), which has been regarded as the de facto standard for privacy-preserving computation.
These approaches mainly focus on adding calibrated noise to gradients or text representations during the training phase so that sensitive user data cannot be inferred from trained LMs.
Nevertheless, they require service providers to collect the original data for LM training.
As such, data owners may still have privacy concerns when service providers are not fully trusted.

To solve the privacy problem from the root, a common paradigm is to let data owners sanitize their data \textit{locally} before releasing them to the service provider, as shown in \Cref{fig:prac}.
Generally, such privatization mechanisms~\cite{Feyisetan2019H, feyisetan2020privacy, yue2021differential} generate a sanitized text document by replacing the original tokens (e.g., characters, words, or $n$-grams) in the original document sequentially with new tokens sampled from output token sets.
Specifically, they adopt the Metric Local Differential Privacy~\cite{chatzikokolakis2013broadening} (MLDP, also known as $d_{\chi}$-privacy), a relaxation of the original DP definition, to provide the privacy and utility guarantees simultaneously.
On the one hand, MLDP inherits the idea of DP to ensure that the outputs of any adjacent input tokens are indistinguishable to protect the original tokens from being inferred.
On the other hand, MLDP also preserves the utility of sanitized texts by assigning higher sampling probabilities to tokens that are semantically closer to the original ones.
In these mechanisms, any metric distance (e.g., Euclidean distance) can be used to measure the semantic similarities between tokens.

However, the above text sanitization mechanisms suffer from two inherent limitations.
First, since MLDP is specific for metric distances satisfying the triangle inequality, they do not apply to non-metric semantic similarity measures in NLP applications such as cosine similarity~\cite{mrksic-etal-2016-counter} and TF-IDF~\cite{salton1988term}.
Second, they cannot achieve good privacy-utility trade-offs, i.e., either having high privacy costs with insufficient protections or resulting in low accuracy of models trained on sanitized data.
We observe that the low accuracy arises as they treat each token in the text equally by assigning each input token with the same output set, which can be excessively large (e.g., the size of the output set is over 80,000).
Such a huge output set leads to high costs for MLDP and thus impedes the model's utility when the privacy budget is tight.

To this end, we propose a novel \underline{Cus}tomized \underline{Text} (CusText) sanitization mechanism that provides more advanced privacy protection at the token level.
Specifically, to generalize CusText to all similarity measures, we turn to a mechanism that satisfies the original $\epsilon$-Differential Privacy ($\epsilon$-DP), i.e., Exponential Mechanism (EM)~\cite{mcsherry2007mechanism}, to sample the output for each input token.
Meanwhile, we inherit the merit of MLDP by designing an appropriate \textit{scoring function} for EM to take into account the semantic similarities between tokens for sampling.
Then, to achieve a better trade-off between privacy and utility, we design a mapping scheme to assign each input token a customized output set of a much smaller size for token-level privacy protection.
Here, we can adjust a customized parameter $K$ that determines the size of the output set for each input token for different utility-privacy trade-offs.
Using the mapping scheme, we exclude most of the tokens that are semantically irrelevant to the input token from consideration and reduce the privacy costs caused by excessive output set sizes.
As the privacy risks of some tokens, e.g., stopwords, are low in practice, we further propose an improved CusText+ mechanism that skips the stopwords in the sampling process to achieve higher utility without incurring greater privacy losses.

Finally, we conduct extensive experiments on three benchmark datasets to demonstrate that CusText achieves better privacy-utility trade-offs than the state-of-the-art text sanitization mechanisms in~\cite{feyisetan2020privacy, yue2021differential}.
More particularly, with the same privacy parameter $\epsilon$, the models trained on texts sanitized by CusText have significantly higher accuracy rates than those sanitized by SANTEXT~\cite{yue2021differential}.
Furthermore, when the utilities of models are comparable, CusText provides better protection against two token inference attacks than SANTEXT.

\section{Related Work}

There have been numerous studies on the vulnerability of deep learning models~\cite{carlini2019secret, song2020information}, including language models~\cite{carlini2021extracting, zhao2022dp} (LMs), against privacy attacks.
In particular, such attacks can recover sensitive user attributes or raw texts from trained models.
Therefore, incorporating privacy mechanisms with rigorous guarantees is vital to protect LMs from privacy attacks.

A few attempts at applying anonymization techniques for i.i.d.~data~\cite{li2007t, machanavajjhala2007diversity} fail to provide strong privacy protection for textual data~\cite{zhao2022dp}.
Then, many efforts~\cite{lyu2020differentially, anil2021large, dupuy2021efficient, hessel2021effective, li2021large,  mireshghallah2021privacy} have been made to preserve the utility of LMs on textual data with provable differential privacy (DP) guarantees.
Following the application of DP in deep learning~\cite{abadi2016deep}, they mainly focus on adding calibrated noise to gradients or text representations during the training phase for both utility and privacy.
However, they need a trustworthy server to collect original texts from data owners for model training and thus cannot be applied to the scenario without trusted servers.

To address privacy issues from the root, different (customized) local differential privacy~\cite{duchi2013local, chatzikokolakis2013broadening} (LDP) mechanisms have been proposed to allow data owners to sanitize their data locally before releasing them to the server.
Due to the high dimensionality and complicated features of textual data, compared with statistical analytics on i.i.d.~data with LDP~\cite{murakami2019utility, ULDP}, it is much more challenging to achieve good utility-privacy trade-offs for LMs with LDP.
To improve the model utility, existing methods~\cite{feyisetan2020privacy, qu2021privacy, yue2021differential} rely on a relaxed notion of metric local differential privacy~\cite{chatzikokolakis2013broadening} (MLDP, also known as $d_{\chi}$-privacy) for text sanitization.
However, they either achieve reasonable accuracy only at a very low privacy protection level (e.g., with a privacy parameter $\epsilon > 10$) or become unusable (around $50\%$ accuracy rate for the benchmark binary classification tasks) with appropriate privacy guarantees (e.g., $\epsilon = 2$).
Thus, there remains much room for improvement in terms of utility-privacy trade-off for differentially private text sanitization, which is the goal of this work.

\section{Preliminaries}
\label{sec: Preliminaries}

Before introducing our CusText mechanism, we briefly review the key concepts, including $\epsilon$-DP and exponential mechanism (EM).

\begin{definition}[$\epsilon$-differential privacy~\cite{dwork2006calibrating}]
For a given privacy parameter $\epsilon \geq 0$, all pairs of adjacent inputs $x, x' \in \mathcal{X}$, and every possible output $y \in \mathcal{Y}$, a randomized mechanism $\mathcal{M}$ is $\epsilon$-differentially private (DP) if it holds that
\begin{equation}
  \frac{\Pr[\mathcal{M}(x)=y]}{\Pr[\mathcal{M}(x')=y]} \leq e^{\epsilon}.
\end{equation}
\end{definition}
By definition, a smaller value of $\epsilon$ corresponds to a higher level of privacy protection.
Conceptually, the notion of $\epsilon$-DP means that an unlimited adversary cannot distinguish the two probabilistic ensembles with sufficiently small $\epsilon$ because the probabilities of adjacent tokens producing the same output token $y$ are similar.
In the context of NLP, we consider any pair of input tokens that share the same output set $\mathcal{Y}$ to be adjacent to each other.
In the rest of this paper, we follow the above definition of adjacent inputs for $\epsilon$-DP.
Next, we define the Exponential Mechanism (EM) commonly used for differentially private item selection from a discrete domain, which naturally fits NLP applications due to the discrete nature of textual data.

\begin{definition}[Exponential Mechanism~\cite{mcsherry2007mechanism}]
\label{em}
For a given scoring function $u: \mathcal{X} \times \mathcal{Y} \to \mathbb{R}$, an exponential mechanism (EM) $\mathcal{M}(\mathcal{X}, u, \mathcal{Y})$ satisfies $\epsilon$-differential privacy if it samples an output token $y\in \mathcal{Y}$ to perturb the input token $x \in \mathcal{X}$ with probability proportional to $e^{\frac{\epsilon \cdot u(x,y)}{2\Delta u}}$, where $u(x,y)$ denotes the score of output token $y$ for input token $x$.
In addition, we use $\Delta u := \max_{y \in \mathcal{Y}} \max_{x,x' \in \mathcal{X}} |u(x,y) - u(x',y)|$ to denote the sensitivity of $u$ for EM.
\end{definition}

From \Cref{em}, we can see that smaller sensitivity makes it harder for adversaries to distinguish the original token from its adjacent tokens.
In practice, for simplicity, we can normalize the scoring function $u$ to scale its sensitivity $\Delta u$ to a specific real number (e.g., $1$).
As such, the sampling probability of each output token $y$ for input token $x$ is only related to $u(x, y)$, as $\epsilon$ and $\Delta u$ are known beforehand, and a larger $u(x,y)$ indicates a higher sampling probability.

In an NLP task, we suppose that each document $D = \langle R_i \rangle_{i=1}^m$ contains $m$ records and each record $R = \langle t_j \rangle_{j=1}^n$ contains $n$ tokens.
We formulate our text sanitization task as follows: Given an input document $D$ containing sensitive information, a set $\mathcal{X}$ of all possible input tokens, a set $\mathcal{Y}$ of all possible output tokens, and a differentially private mechanism $\mathcal{M}$ (e.g., EM in this work), it performs the mechanism $\mathcal{M}$ on each input token $t_j \in D$ to replace it with an output token $t'_j$ from $\mathcal{Y}$ if $t_j \in \mathcal{X}$.
All the tokens after replacement form the sanitized document, i.e., $D'= \langle R'_i \rangle_{i=1}^m$ and $R' = \langle t'_j \rangle_{j=1}^n$.

Following the prior work on text sanitization~\cite{qu2021privacy, feyisetan2020privacy, yue2021differential}, we consider a \textit{semi-honest threat model} under the LDP setting where data owners (e.g., individuals or institutions) only submit their sanitized documents to the service provider.
Malicious service providers may try to infer sensitive information from their received data.
We assume that adversaries only have access to sanitized texts, and all algorithms and mechanisms are publicly known.
Moreover, adversaries have unlimited computation resources.

\section{The CusText Mechanism}

\begin{figure}[t]
\centering
\includegraphics[width=\linewidth]{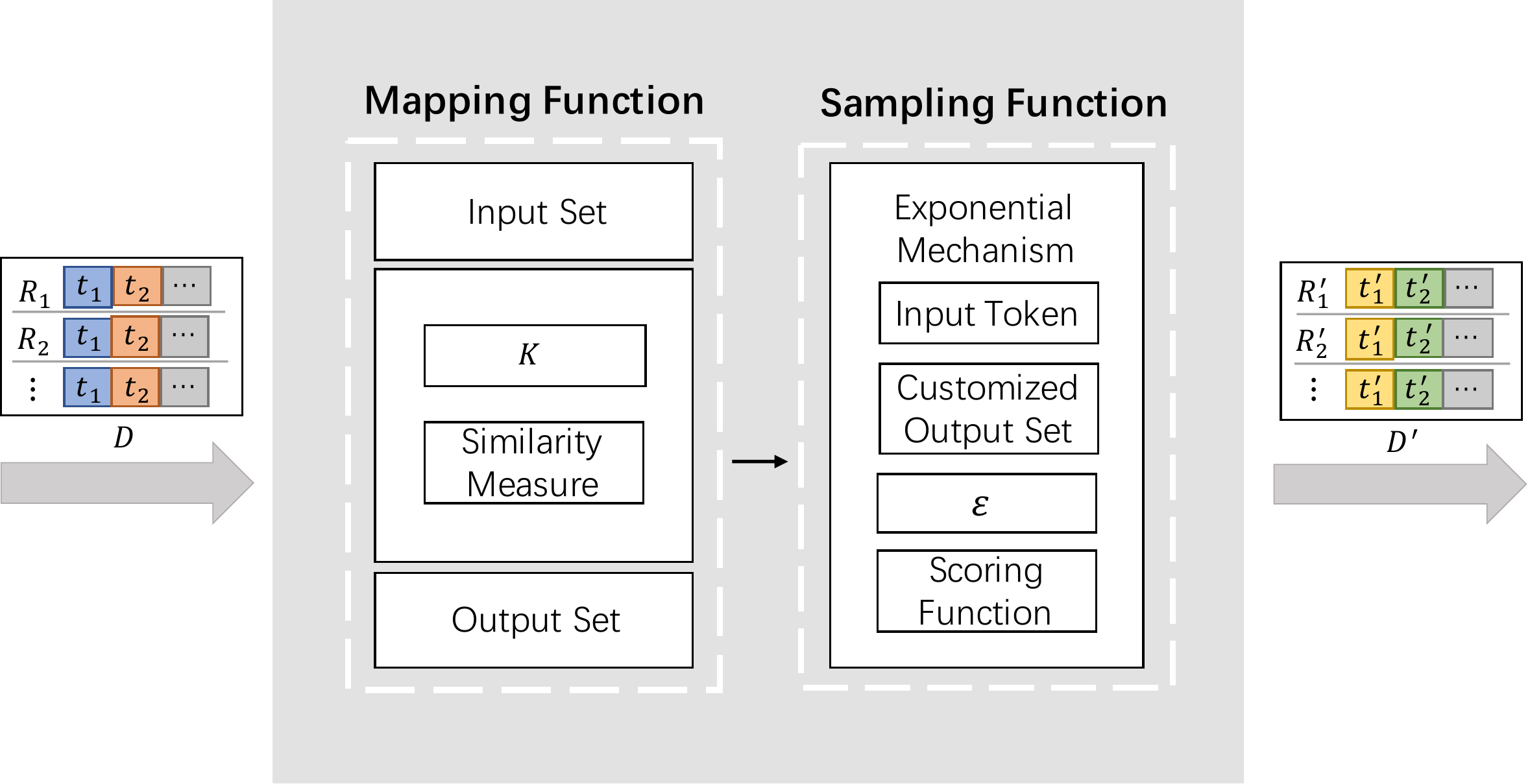}
\caption{An overview of the CusText method.}
\label{fig:process}
\end{figure}

\begin{figure}[t]
\centering
\includegraphics[width=\linewidth]{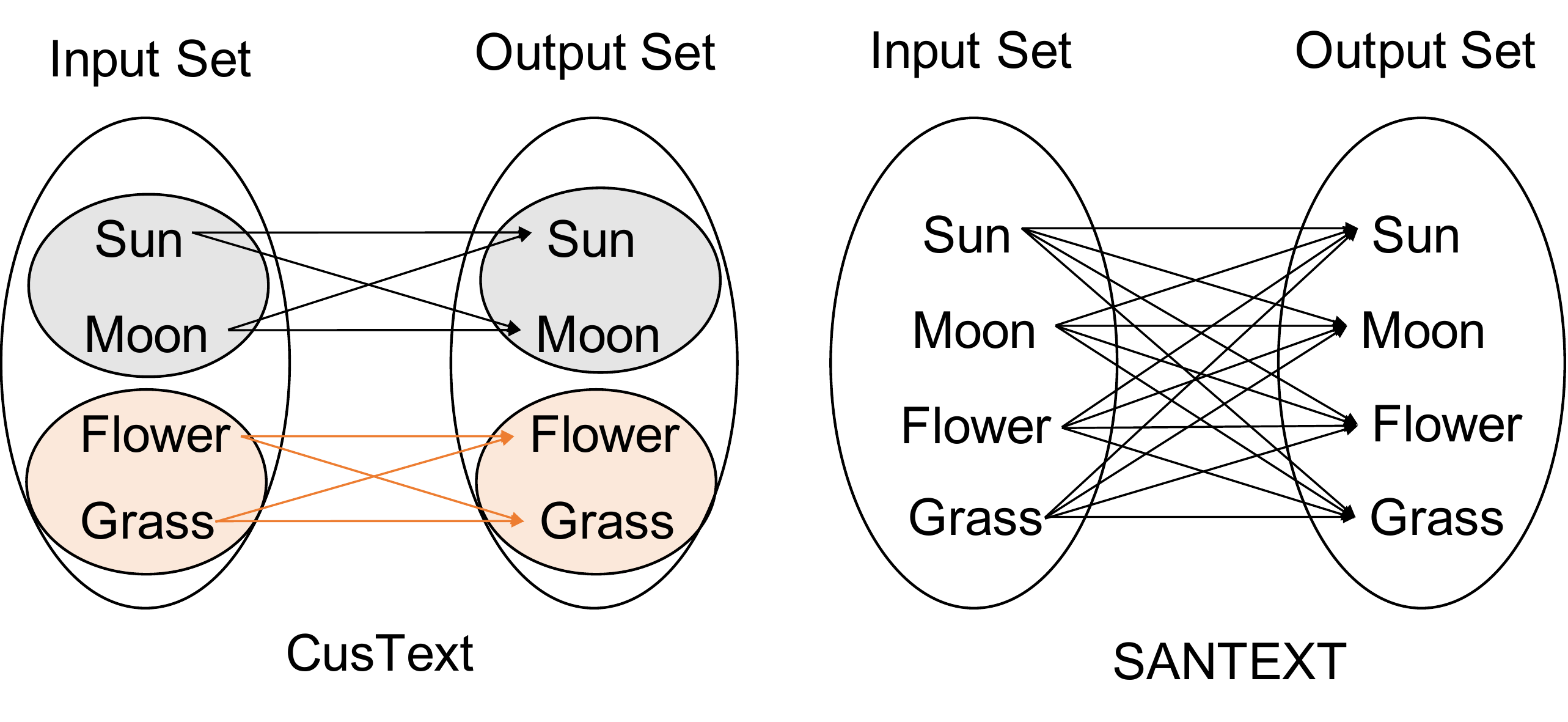}
\caption{A comparison of the mapping schemes of SANTEXT and CusText.}
\label{fig:com}
\end{figure}

An overview of our customized text (CusText) sanitization mechanism is presented in \Cref{fig:process}.
In general, it replaces each token in the original text document with a new token to achieve the privacy guarantee.
It consists of two components:
(1) a mapping function $f_\mathsf{map}: \mathcal{X} \to \{\mathcal{Y}^\prime \subseteq \mathcal{Y}\}$ that determines the output set $\mathcal{Y}^\prime_{j}$ for each input token $x_j \in \mathcal{X}$ based on semantic relevance;
(2) a sampling function\footnote{For any $\mathcal{Y'} \subseteq \mathcal{Y}$, $\mathcal{X'} = \{x \in \mathcal{X} \,|\, f_\mathsf{map}(x) = \mathcal{Y'}\}$.} $f_\mathsf{sample}: \mathcal{X'} \to \mathcal{Y'}$ based on the exponential mechanism to sample a new token from an output set to sanitize the input token.
Specifically, our CusText mechanism first obtains the output set $\mathcal{Y}^\prime_j$ for each $t_j \in D$ according to $f_\mathsf{map}$, i.e., $ \mathcal{Y}^\prime_j = f_\mathsf{map}(t_j)$, then samples an output token $t'_j$ from $\mathcal{Y}^\prime_{j}$ according to $f_\mathsf{sample}$, i.e., $ t'_j = f_\mathsf{sample} (t_j)$. Finally, after applying CusText on each input token $t_j$ in $D$, the sanitized document $D'$ is formed by all output tokens.

\subsection{Mapping Function}
\label{sec:mr}

In our CusText mechanism, the mapping function $f_\mathsf{map}: \mathcal{X} \to \{\mathcal{Y}^\prime \subseteq \mathcal{Y}\}$ decides the output set for each input token.
If a bunch of input tokens in $\mathcal{X}$ are mapped to the same output set $\mathcal{Y'}$, we say that they belong to the same input set $\mathcal{X'} \subseteq \mathcal{X}$ and are adjacent to each other.
For the SANTEXT mechanism~\cite{yue2021differential}, the function $f_\mathsf{map}: \mathcal{X} \to \mathcal{Y}$ simply maps every input token $x \in \mathcal{X}$ to all tokens in the output set $\mathcal{Y}$.
Since the size of the output set is excessively large in SANTEXT, the chances that the output token is semantically irrelevant to the original token become higher if the privacy budget is tight, thus leading to poor model utility.
To overcome the above problem, CusText customizes the output set of each input token.
A comparison of the mapping schemes of CusText and SANTEXT is shown in \Cref{fig:com}.
Before introducing how to construct $f_\mathsf{map}$, we first discuss the requirements for mapping generation.

\noindent\textbf{Mapping Strategy.}
According to the sizes of $\mathcal{X'}$ and $\mathcal{Y'}$ as indicated by the mapping function $f_\mathsf{map}$, we can categorize the token mappings into four types: $1$-to-$1$, $N$-to-$1$, $1$-to-$N$, and $N$-to-$M$, where $1$, $N$, and $M$ denote the size of the input/output token sets and $N, M > 1$.
Theoretically, CusText can provide $\epsilon$-differential privacy protection to all input tokens only if the mappings of all input tokens in the set $\mathcal{X}$ are $N$-to-$M$ or $N$-to-$1$ mappings so that every input token in $\mathcal{X}$ has at least one adjacent token.
This is because the goal of applying $\epsilon$-DP is to make any two adjacent tokens indistinguishable so that the input token cannot be effectively inferred.
Moreover, following prior work~\cite{feyisetan2020privacy, yue2021differential}, we consider that $\mathcal{X}$ is equal to $\mathcal{Y}$ (i.e., $\mathcal{X} = \mathcal{Y}$) in CusText, as they both correspond to the vocabulary of a specific language.
Also, any input token $x$ is always included in its output set because it must be the closest to itself.
Next, we describe our mapping generation that can satisfy all the above requirements.

\begin{algorithm}[t]
  \small
  \caption{Token Mapping Generation}
  \label{alg:1}
  \renewcommand{\algorithmicrequire}{\textbf{Input:}}
  \renewcommand{\algorithmicensure}{\textbf{Output:}}
  \begin{algorithmic}[1]
    \REQUIRE Customization parameter $K$, input set $\mathcal{X}$, output set $\mathcal{Y} = \mathcal{X}$, similarity measure $d$
    \ENSURE Mapping Function $f_\mathsf{map}$
    \WHILE {$|\mathcal{X}| \geq K$}
      \STATE Pick an arbitrary token $x$ from $\mathcal{X}$ \label{ln:map:s}
      \STATE Initialize an output set $\mathcal{Y'} = \{ x \}$ for $x$
      \FORALL {$y \in \mathcal{Y} \setminus \{ x \}$}
        \STATE Compute the similarity $d(x,y)$ of $x$ and $y$
      \ENDFOR
      \STATE Add the top-$(K - 1)$ tokens that are semantically closest to $x$ to $\mathcal{Y'}$ based on $d(\cdot, \cdot)$
      \FORALL {$x' \in \mathcal{Y}'$}
        \STATE Assign the output set of $x'$ as $\mathcal{Y'}$
      \ENDFOR
      \STATE Update $\mathcal{X} \gets \mathcal{X} \setminus \mathcal{Y'}$ and $\mathcal{Y} \gets \mathcal{Y} \setminus \mathcal{Y'}$ \label{ln:map:t}
    \ENDWHILE
    \STATE Perform Lines~\ref{ln:map:s}--\ref{ln:map:t} for the remaining tokens in $\mathcal{X}$ and $\mathcal{Y}$ with customization parameter $K' = |\mathcal{X}|$
    \RETURN $f_\mathsf{map}$
  \end{algorithmic}
\end{algorithm}

\smallskip
\noindent\textbf{Mapping Function Generation.}
The generation of the mapping function $f_\mathsf{map}: \mathcal{X} \to \{ \mathcal{Y}^\prime \subseteq \mathcal{Y} \}$ is to assign the customized output set for each input token based on semantic relevance.
The semantic relevance can be defined by any similarity measure $d: \mathcal{X} \times \mathcal{Y} \to \mathbb{R}$.
In practice, we use the Euclidean distance or cosine similarity on the vector representations of tokens, such as Word2Vec~\cite{mikolov2013efficient},  GloVe~\cite{pennington2014glove}, and Counter-Fitting~\cite{mrksic-etal-2016-counter} as the similarity measure.
Then, we fix the sizes of all output sets to $K$.
Specifically, we pick an arbitrary unmapped token $x \in \mathcal{X}$, find the $K$ tokens semantically closest to $x$, generate an $K$-to-$K$ mapping from all the $K$ tokens to themselves, and remove the mapped tokens from $\mathcal{X}$ and $\mathcal{Y}$ at each round until either all tokens are mapped or fewer than $K$ tokens remain unmapped.
In the latter case, the remaining tokens will constitute a $K'$-to-$K'$ mapping where $K' \in [1, K)$.
The pseudocode of generating the mapping function $f_\mathsf{map}$ is presented in \Cref{alg:1}.

\subsection{Sampling Function}

Based on the mapping function $f_\mathsf{map}: \mathcal{X} \to \{ \mathcal{Y}^\prime \subseteq \mathcal{Y} \}$, a sampling function $f_\mathsf{sample}: \mathcal{X'} \to \mathcal{Y'}$ is designed to sample the output token for each input token.
CusText adopts the exponential mechanism~\cite{mcsherry2007mechanism} (EM) for sampling.
We need to design an appropriate scoring function for EM to strike a good utility-privacy trade-off.
We obey the following two rules when designing the scoring function $u: \mathcal{X'} \times \mathcal{Y'} \to \mathbb{R}$.
\begin{enumerate}
  \item The score of each pair of input and output tokens should be bounded, i.e., $\forall x \in \mathcal{X'}$, $\forall y \in \mathcal{Y'}$, $u(x,y) < B$, so that the sensitivity $\Delta u$ of $u$ is bounded for satisfying $\epsilon$-DP.
  \item The higher the semantic similarity between a pair of input and output tokens is, the higher the score is, i.e., $\forall x \in \mathcal{X'}$, $\forall y, y' \in \mathcal{Y'}$, if $y$ is semantically closer to $x$ than $y'$, $u(x, y) > u(x, y')$. This ensures the candidates semantically closer to $x$ have higher probabilities of being sampled, which inherits the advantage of $d_{\chi}$-privacy~\cite{chatzikokolakis2013broadening}.
\end{enumerate}

For the scoring function, we are based on the same similarity function as used in the mapping scheme, e.g., Euclidean distance or cosine similarity on the vector representations of tokens~\cite{mikolov2013efficient, pennington2014glove, mrksic-etal-2016-counter}.
Generally, according to the correlation between scores and semantic closeness, all the similarity measures can be categorized into two types, i.e., \textit{negative correlation} and \textit{positive correlation}.
For instance, Euclidean distance and cosine similarity are \textit{negative} and \textit{positive correlation} measures, respectively, as a smaller Euclidean distance and a larger cosine value between two vectors imply higher semantic closeness of their corresponding tokens.
Next, we will design scoring functions for both types of similarity measures.

\smallskip
\noindent\textbf{Scoring Function for Negative Correlation Measures.}
We take Euclidean distance as an example to design the scoring function $u: \mathcal{X'} \times \mathcal{Y'} \to \mathbb{R}$.
For any input set $\mathcal{X'}$ and its corresponding output set $\mathcal{Y'}$, we first compute the Euclidean distance $d(x, y)$ between each $x \in \mathcal{X'}$ and $y \in \mathcal{Y'}$.
Specifically, we have $d(x, y) = \| \Phi(x) - \Phi(y) \|_2$, where $\Phi(x)$ and $\Phi(y)$ are the vector representations of $x$ and $y$, respectively.
Then, we normalize the distances of all pairs of tokens to the range $[0,1]$ as $d'(x, y) = \frac{d(x, y) - d_{min}}{d_{max} - d_{min}}$,
where $d_{min} = \min_{x \in \mathcal{X'}, y \in \mathcal{Y'}} d(x, y)$ and $d_{max} = \max_{x \in \mathcal{X'}, y \in \mathcal{Y'}} d(x, y)$.
Finally, we transform the normalized distance $d'(x, y)$ into the score of output token $y$ for input token $x$ as $u(x, y) = - d'(x, y)$.
After the above transformation, a more similar pair $x, y$ of input and output tokens has a higher score $u(x, y)$.
Finally, by repeating the above steps on all disjoint partitions of adjacent tokens with the same $\mathcal{X'}$ and $\mathcal{Y'}$, we have obtained the scoring functions for all tokens.

\smallskip
\noindent\textbf{Scoring Function for Positive Correlation Measures.}
We take cosine similarity as another example to design the scoring function $u$.
For any input set $\mathcal{X'}$ and its corresponding output set $\mathcal{Y'}$, we also compute the cosine similarity $\cos(x, y)$ between each $x \in \mathcal{X'}$ and $y \in \mathcal{Y'}$, where $\cos(x, y) = \frac{\langle \Phi(x), \Phi(y) \rangle}{\| \Phi(x) \| \cdot \| \Phi(y) \|}$ and $\Phi(x)$ and $\Phi(y)$ are the vector representations of $x$ and $y$.
Then, the normalization procedure is the same as that for Euclidean distance, but we use the normalized distance, instead of its additive inverse, in the score function, i.e., $u(x, y) = \frac{d(x, y) - d_{min}}{d_{max} - d_{min}}$.
Finally, we repeat the above steps on all disjoint partitions of adjacent tokens to obtain all scoring functions.

\begin{algorithm}[t]
  \small
  \caption{Document Sanitization}
  \label{alg:2}
  \renewcommand{\algorithmicrequire}{\textbf{Input:}}
  \renewcommand{\algorithmicensure}{\textbf{Output:}}
  \begin{algorithmic}[1]
    \REQUIRE Original document $D = \left \langle R_i \right \rangle_{i=1}^m$, sampling function $f_\mathsf{sample}$, stopword list $T$
    \ENSURE Sanitized document $D'$
    \STATE Initialize the sanitized document $D' = \emptyset$
    \FORALL {record $R \in D$}
      \STATE Initialize the sanitized record $R' = \emptyset$
      \FORALL {token $x \in R$}
        \IF{`CusText+' is used and $x \in T$}
          \STATE Append $x$ to $R'$
        \ELSE
          \STATE $x' \gets f_\mathsf{sample} (x)$ and append $x$ to $R'$
        \ENDIF
      \ENDFOR
      \STATE Add $R'$ to $D'$
    \ENDFOR
    \RETURN $D'$
  \end{algorithmic}
\end{algorithm}

\smallskip
\noindent\textbf{Sampling Procedure.}
After acquiring the scoring function $u$ for each input token $x$, the sampling function $f_\mathsf{sample}$ is used to generate the sanitized token $x'$ for $x$ based on the exponential mechanism $\mathcal{M}(\{x\}, u, \mathcal{Y'})$ with a privacy parameter $\epsilon > 0$.
The pseudocode of sanitizing a document based on $f_\mathsf{sample}$ is provided in \Cref{alg:2}.
Theoretically, it guarantees that $f_\mathsf{sample}$ satisfies $\epsilon$-DP.
For any input set $\mathcal{X'}$ and its corresponding output set $\mathcal{Y'}$, 
the sensitivity $\Delta u$ between any two adjacent input tokens $x,x' \in \mathcal{X'}$ is bound by $1$ according to the design of the scoring function $u$, i.e.,
\begin{equation*}
  \Delta u= \max \limits_{y \in \mathcal{Y'}} \max \limits_{x,x' \in \mathcal{X'}} | u(x,y) - u(x',y) | = 1
\end{equation*}
Given a privacy parameter $\epsilon > 0$, the probability of obtaining an output token $y \in \mathcal{Y'}$ for an input token $x \in \mathcal{X'}$ is as follows:
\begin{equation*}
  \Pr[f_\mathsf{sample}(x) = y] = \frac{\exp(\frac{\epsilon u(x, y)}{2 \Delta u})}{\sum_{y' \in \mathcal{Y'}} \exp(\frac{\epsilon u(x, y')}{2 \Delta u})}
\end{equation*}
We can prove that the sampling function $f_\mathsf{sample}$ satisfies $\epsilon$-DP because, for any two input tokens $x, x' \in \mathcal{X'}$ and output token $y \in \mathcal{Y'}$, it holds that
\begin{align*}
&\frac{\Pr[f_\mathsf{sample}(x) = y]}{\Pr[f_\mathsf{sample}(x') = y]} = \frac{\frac{\exp(\frac{\epsilon u(x, y)}{2 \Delta u})}{\sum_{y' \in \mathcal{Y'}} \exp(\frac{\epsilon u(x, y')}{2 \Delta u})}}{\frac{\exp(\frac{\epsilon u(x',y)}{2 \Delta u})}{\sum_{y' \in \mathcal{Y'}} \exp(\frac{\epsilon u(x',y') }{2 \Delta u})}} \\
& = e^{\frac{\epsilon \cdot (u(x,y) - u(x',y))}{2 \Delta u}} \cdot \Big(\frac{\sum_{y' \in \mathcal{Y'}} \exp(\frac{\epsilon u(x',y') }{2 \Delta u})}{\sum_{y' \in \mathcal{Y'}}\exp(\frac{\epsilon u(x,y') }{2 \Delta u})}\Big) \\
& \leq e^{\frac{\epsilon}{2}} \cdot e^{\frac{\epsilon}{2}} \cdot \Big(\frac{\sum_{y' \in \mathcal{Y'}} \exp(\frac{\epsilon u(x, y')}{2 \Delta u})}{\sum_{y' \in \mathcal{Y'}} \exp(\frac{\epsilon u(x,y') }{2 \Delta u})}\Big) = e^{\epsilon}.
\end{align*}

\subsection{The CusText+ Mechanism}

Since not all tokens contain sensitive information, our CusText mechanism that replaces all tokens might be over-protective.
Therefore, we can retain non-sensitive original tokens with low privacy risk (e.g., stopwords) to improve the utility of the sanitized text.
In practice, we have a predefined list of stopwords $T$ (e.g., the collection of stopwords in the NLTK library), check whether each token $x$ is included in $T$, and keep $x$ in the sanitized document if $x \in T$ or replace $x$ with $x' = f_\mathsf{sample}(x)$ otherwise.
The above procedure is called the CusText+ mechanism and is also described in \Cref{alg:2}.


\begin{table*}[!ht]
  \centering
  \small
  \begin{tabular}{|c|c|c|c|c|c|c|c|c|c|}
  \hline
  \multirow{2}{*}{\textbf{Mechanisms}} & \multicolumn{3}{c|}{\textbf{SST2}} & \multicolumn{3}{c|}{\textbf{MedSTS}} &\multicolumn{3}{c|}{\textbf{QNLI}} \\ \cline{2-10}
  ~ & $\epsilon  = 1$ & $\epsilon = 2$ & $\epsilon = 3$ & $\epsilon = 1$ & $\epsilon = 2$ & $\epsilon = 3$ & $\epsilon = 1$ & $\epsilon = 2$ & $\epsilon = 3$ \\ \hline
  Random & \multicolumn{3}{c|}{0.5014} & \multicolumn{3}{c|}{0.0382} &\multicolumn{3}{c|}{0.5037} \\ \cline{1-10}
  FBDD & 0.5022 & 0.5041 & 0.5032 & 0.0321 & 0.0368 & 0.0411 & 0.5021 & 0.5152  & 0.5368 \\ 
  SANTEXT & 0.5014 & 0.4827 & 0.5091 & 0.0850 & 0.1673 & 0.1124 & 0.5304 & 0.5302 & 0.5357 \\ 
  CusText & 0.6985 & 0.7172 & 0.7029 & 0.4957 & 0.5112 & 0.5242 & 0.6926 & 0.6884 & 0.7133 \\
  SANTEXT+  & 0.7211 & 0.7446 & 0.7260 & 0.4143 & 0.4271 & 0.5423 & 0.7607 & 0.7636 & 0.7493 \\  
  CusText+ & 0.7501 & 0.7452 & 0.7683 & 0.6172 & 0.6316 & 0.6213  & 0.7528  & 0.7602  & 0.7740 \\
  \hline
  Original & \multicolumn{3}{c|}{0.9050} & \multicolumn{3}{c|}{0.7598} &\multicolumn{3}{c|}{0.9096} \\
  \hline
  \end{tabular}
  \caption{Utility comparison of different sanitization mechanisms at similar privacy levels.}
  \label{tb:main-exp}
\end{table*}

\section{Experiments}

\subsection{Experimental Setup}

Following~\cite{feyisetan2020privacy, yue2021differential}, we choose two datasets from the GLUE benchmark~\cite{wang2018glue} and one medical dataset MedSTS~\cite{wang2020medsts}, which all contain sensitive information, in our experiments. Detailed descriptions of the three datasets are as follows:
\begin{itemize}
  \item \textbf{SST-2} is a popular movie reviews dataset with 67k training samples and 1.8k test samples for sentiment classification, where \textit{accuracy} is used as the evaluation metric.
  \item \textbf{MedSTS} is a medical dataset with 1,642 training samples and 412 test samples for semantic similarity computation, where \textit{Pearson correlation coefficient} is used for evaluation.
  \item \textbf{QNLI} is a sentence dataset with 105k training samples and 5.2k test samples for sentence-pair classification, where \textit{accuracy} is used as the evaluation metric.
\end{itemize}

In the experiments, we compare CusText with two existing text sanitization mechanisms, i.e., FBDD~\cite{feyisetan2020privacy} and SANTEXT~\cite{yue2021differential}.
In the training phase, we perform each mechanism to sanitize the training data and then use the sanitized documents to fine-tune the pre-trained model.
In the evaluation phase, we sanitize the test data by the same mechanism as used for training.
When producing the sanitized documents, both the input set $\mathcal{X}$ and output set $\mathcal{Y}$ are assigned to the vocabulary of Counter-Fitting~\cite{mrksic-etal-2016-counter} (of size 65,713), and out-of-vocabulary (OOV) tokens except numbers are retained.
For a fair comparison, we adopt the same vocabulary in GloVe~\cite{pennington2014glove} as in Counter-Fitting.
The Euclidean distance and cosine similarity are used as the similarity measures for GloVe and Counter-Fitting, respectively.
We use the stopword list in NLTK for CusText+.
For each downstream task, we set the maximum sequence length to $128$ and the training epoch to $3$.
On the SST2 and QNLI datasets, we set the batch size to $64$ and the learning rate to $2 \times 10^{-5}$ using \textit{bert-base-uncased}\footnote{\url{https://huggingface.co/bert-base-uncased}} as the pre-trained model.
On the MedSTS dataset, we set the batch size to $8$ and the learning rate to $5 \times 10^{-5}$ using \textit{ClinicalBERT}~\cite{alsentzer2019publicly} as the pre-trained model.
Other hyper-parameters are the same as those used in the default Transformer model~\cite{2020Transformers}.
All experiments were conducted on a server with two Intel Xeon Silver 4210R 2.40GHz CPUs and one NVIDIA Tesla V100 SXM2 (32GB).

\subsection{Experimental Results}

\noindent\textbf{Comparison of Different Mechanisms for Text Sanitization.}
In this experiment, we fix the customization parameter $K$ to $20$ in CusText and CusText+ and vary the privacy parameter $\epsilon = 1, 2, 3$ for DP.
The evaluation of the effect of $K$ on the performance of CusText will be presented later.
Furthermore, we choose GloVe as the token embedding in CusText and CusText+ for a fair comparison since FBDD, SANTEXT, and SANTEXT+ cannot apply the Counter-Fitting embedding.
This is because they only work with metric distances (e.g., Euclidean distance in GloVe) due to the inherent limitation of MLDP and thus cannot handle the non-metric cosine similarity in Counter-Fitting.
Finally, because a mechanism will be $\epsilon$-DP if it is $\epsilon'$-MLDP~\cite{chatzikokolakis2013broadening}, where $\epsilon = \epsilon' \cdot d_{max}$ and $d_{max} = \max_{x \in \mathcal{X}, y \in \mathcal{Y}} d(x, y)$, we re-scale the privacy parameter $\epsilon$ in FBDD, SANTEXT, and SANTEXT+ with $d_{max}$ to align their privacy levels to be similar to our mechanisms.

Table~\ref{tb:main-exp} presents the utilities of different text sanitization mechanisms with $\epsilon$-DP ($\epsilon = 1, 2, 3$) on three datasets.
The results demonstrate the huge advantages of CusText compared with two existing mechanisms, i.e., FBDD and SANTEXT, which achieves over 20\% improvements in accuracy on the SST-2 and QNLI datasets and more than 50\% improvement in Pearson correlation coefficient on the MedSTS dataset.
Compared with SANTEXT and CusText, their improved versions, i.e., SANTEXT+ and CusText+, exhibit significantly better performance because they keep some original tokens to preserve original semantics.
Generally, the results indicate the superior performance of CusText by showing that using a customized, smaller output set for each input token can lead to better utilities at similar (theoretical) privacy levels.

\begin{table*}[t]
  \small
  \centering
  \begin{tabular}{|c|c|c|c|c|c|c|c|c|c|c|c|}
  \hline
  \multirow{2}{*}{\textbf{Token}} & \multicolumn{3}{c|}{SANTEXT} & \multicolumn{4}{c|}{CusText (GloVe)} & \multicolumn{4}{c|}{CusText (Counter-Fitting)} \\
  \cline{2-12}
  & $\epsilon'=1$ & $\epsilon'=2$ & $\epsilon'=3$ & $\epsilon=1$ & $\epsilon=2$ & $\epsilon=3$ & $\epsilon=8$ & $\epsilon=1$ & $\epsilon=2$ & $\epsilon=3$ & $\epsilon=8$ \\
  \hline
  \textsf{she} & 2350 & 35 & 4 & 1000 & 200 & 80 & 5 & 5500 & 1000 & 320 & 4 \\
  \textsf{car} & 1300 & 14 & 1 & 1220 & 250 & 90 & 6 & 420000 & 90000 & 31000 & 3200 \\
  \textsf{alice} & 1550 & 20 & 3 & 1190 & 240 & 100 & 6 & 1700 & 360 & 120 & 9 \\
  \textsf{happy} & 3200 & 55 & 4 & 1490 & 290 & 110 & 8 & 320000 & 55000 & 21500 & 1500 \\
  \hline
  \textbf{Accuracy} & 0.4959 & 0.5799 & 0.7958 & 0.6985 & 0.7172 & 0.7029 & 0.8155 & 0.7117 & 0.7370 & 0.7298 & 0.7957 \\
  \hline
  \end{tabular}
  \caption{Results for query attacks on four selected tokens in the SST-2 dataset.}
  \label{tb:query}
\end{table*}

\smallskip
\noindent\textbf{Privacy-Utility Trade-off.}
Subsequently, we compare SANTEXT and CusText in terms of privacy-utility trade-offs.
As shown in~\cite{yue2021differential} as well as our previous results, FBDD has lower performance than SANTEXT and CusText and thus is not compared in the remaining experiments anymore.
To alleviate the effects of different DP definitions in SANTEXT and CusText, we do not use the privacy parameter $\epsilon$, which corresponds to the worst possible privacy leakage but may not reveal the privacy protection level in practice.
Alternatively, we adopt two privacy attacks to evaluate the privacy protection levels: One is the \emph{Mask Token Inference Attack} in \cite{yue2021differential}, and the other is \emph{Query Attack} proposed in this work.

\begin{figure}[t]
  \centering
  \includegraphics[width=0.9\linewidth]{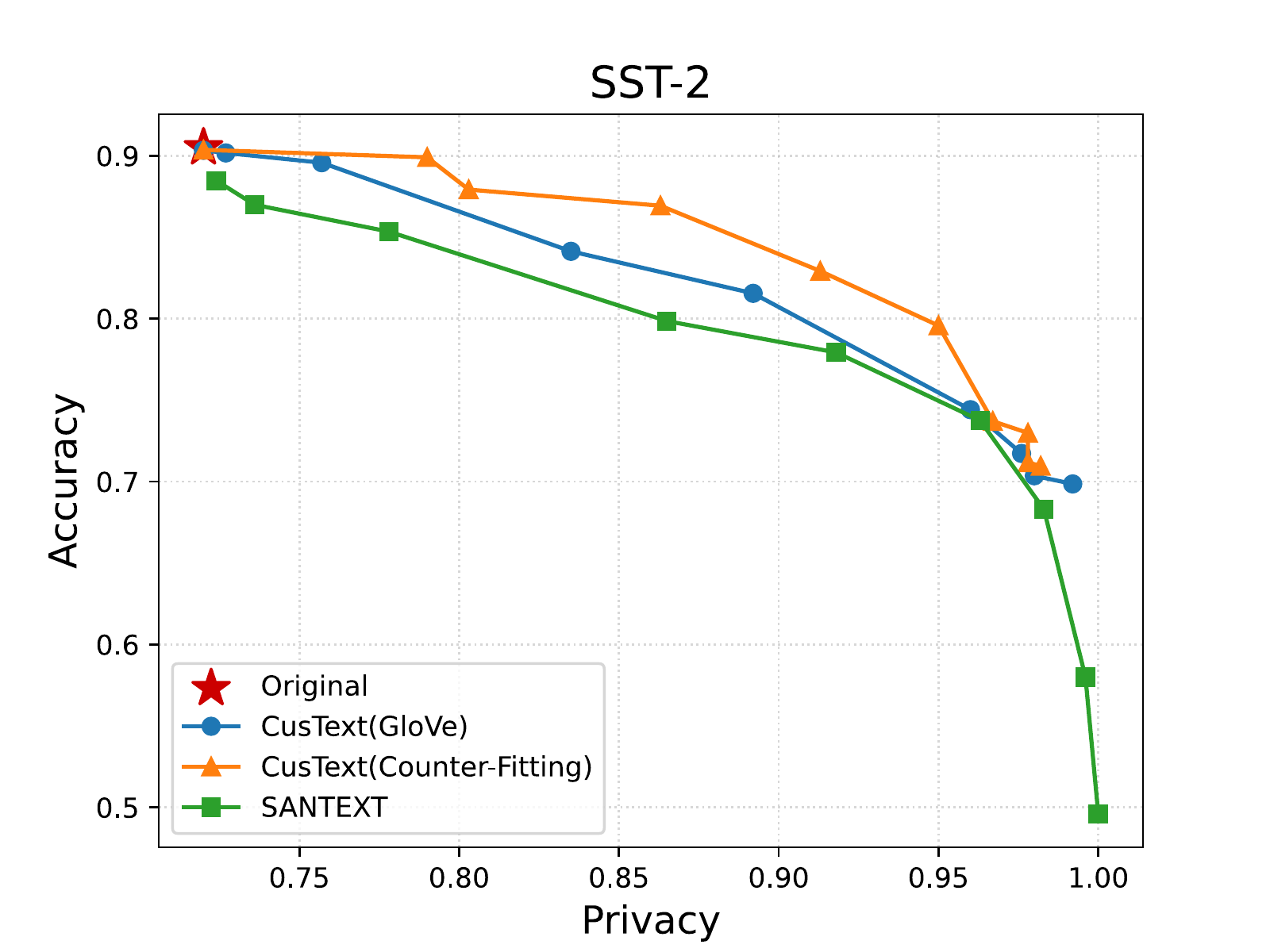}
  \caption{Privacy-utility trade-offs in terms of success rates of mask token inference attacks vs.~accuracy rates by varying the privacy parameter $\epsilon \in [0.01, 50]$ on the SST-2 dataset. Here, ``Original'' denotes the result on unsanitized data.}
  \label{fig:mask-sst2}
\end{figure}

We first present the results for mask token inference attacks.
To recover raw texts from sanitized texts, an adversary can use the pre-trained BERT model to help infer the original tokens since it is trained via masked language modeling.
It replaces each token with a special token ``\textsf{[MASK]}'' in the sanitized text sequentially, inputs the masked text to BERT, and acquires the predicted output of ``\textsf{[MASK]}'' as the original token.
Then, we consider the attack successful if the output token is the same as the input.
Finally, we compute the success rate among all attacks, denoted as $r_{mask}$, and define the privacy protection level as $1 - r_{mask}$.

Figure~\ref{fig:mask-sst2} illustrates the privacy-utility trade-offs of CusText (based on GloVe and Counter-Fitting, respectively) and SANTEXT (based on GloVe) by varying the value of $\epsilon$ on the SST-2 dataset.
The results confirm that CusText achieves better utility-privacy trade-offs than SANTEXT and remains a relatively good utility (accuracy at around $0.7$) when the privacy level approaches $1$ (over $0.98$). In comparison, SANTEXT degenerates to a random classifier (accuracy at around $0.5$).
Meanwhile, the results also imply that Counter-Fitting works better with CusText than GloVe.
The higher performance of Counter-Fitting can be attributed to its better representations of synonyms.

We then describe the results for query attacks.
Since the input token is contained in its corresponding output set and always has the highest score, the probability that it is sampled by $f_\mathsf{sample}$ is also the highest among all output tokens.
An adversary can determine the input token by querying the data owner for the sanitized document multiple times, as the input token will have the highest frequency among all output tokens after a sufficiently large number of queries.
Thus, we use the smallest number $N$ of queries an adversary needs to infer the input token at a confidence level of $95\%$ as a new measure of the privacy protection level.
Here, the larger the value of $N$ is, the higher the level of privacy protection is.
In the experiment, we obtain the value of $N$ by using the Monte Carlo method~\citep{gentle2009monte} to sample the output tokens until the confidence level of determining the input token from the output distribution reaches $95\%$.

\Cref{tb:query} further confirms that CusText achieves better privacy-utility trade-offs than SANTEXT.
Although SANTEXT achieves a good utility when $\epsilon' = 3$ (i.e., with $3$-MLDP), it almost provides no privacy protection as input tokens can be inferred by performing only a few queries.
CusText (with either GloVe or Counter-Fitting) remains relatively good privacy protection levels when $\epsilon = 3$ (i.e., with $3$-DP) while achieving high utilities.
Generally, Counter-Fitting also outperforms GloVe for CusText.
But the privacy protections for different tokens vary very much for Counter-Fitting: ``\textsf{she}'' and ``\textsf{alice}'' are more vulnerable than ``\textsf{car}'' and ``\textsf{happy}''. This is because ``\textsf{she}'' and ``\textsf{alice}'' are mapped with semantically less relevant tokens than themselves in the mapping function generation.

\begin{figure}[t]
  \centering
  \includegraphics[width=0.9\linewidth]{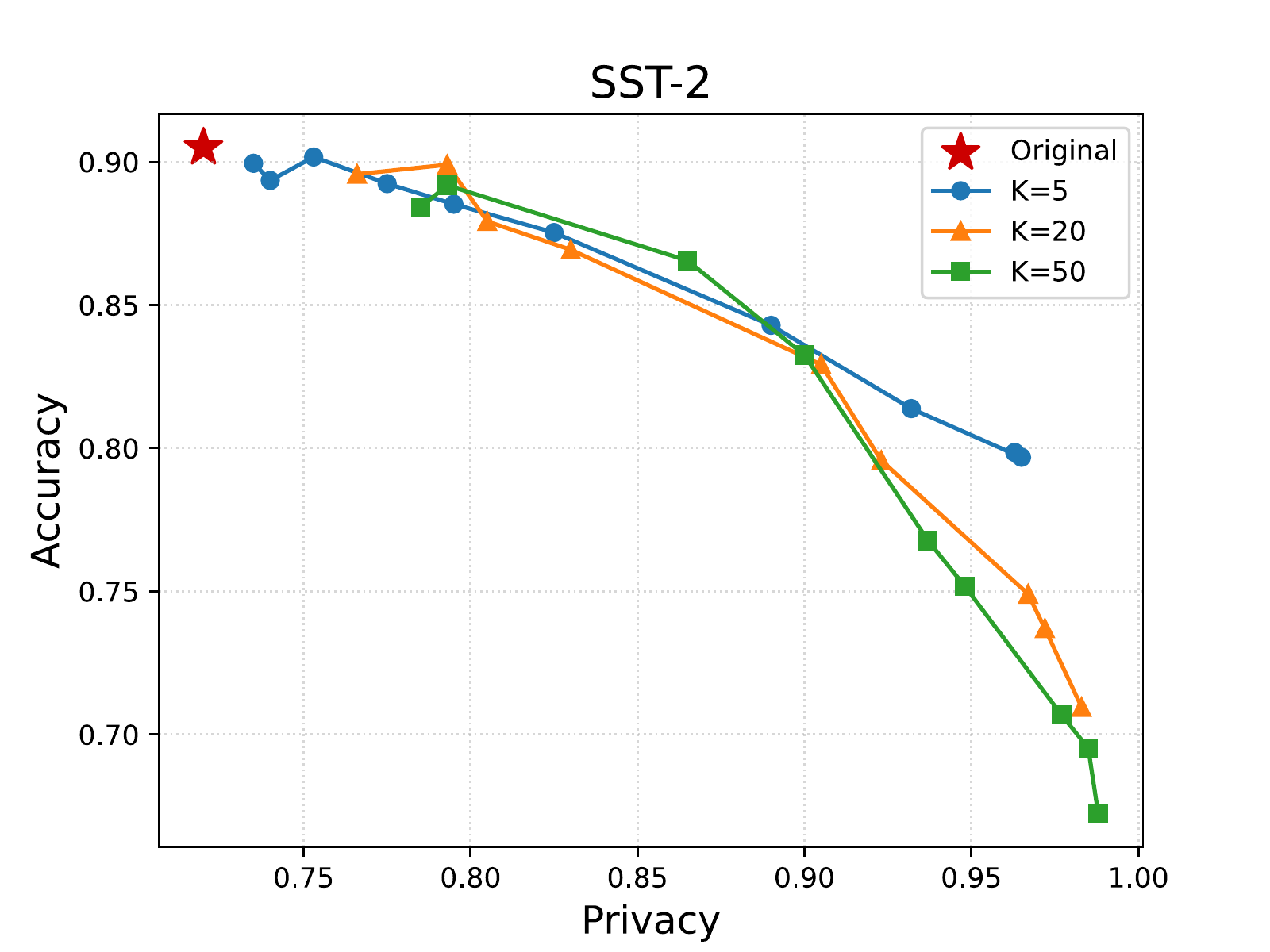}
  \caption{Privacy-utility trade-offs of CusText with different customization parameters $K$ by varying the privacy parameter $\epsilon \in [0.001, 50]$ on the SST-2 dataset.
  }
  \label{fig:k}
\end{figure}

\smallskip
\noindent\textbf{Effect of $K$ on CusText.}
To test the effect of $K$ on CusText in practice, we study the privacy-utility trade-offs with different customization parameters $K = 5, 20, 50$ on the SST-2 dataset.
We choose the mask token inference attack as the privacy metric since its performance is more semantically related.
Then, we use Counter-Fitting for its better performance than GloVe, as depicted previously.

The results for different $K$'s are presented in \Cref{fig:k}.
We observe that the performance of CusText is generally stable for different $K$'s.
But it achieves slightly better utilities when $K$ is smaller at relatively higher privacy protection levels ($>0.9$).
This is because, on the one hand, the semantic similarity of output tokens to the input token will be higher when $K$ is smaller.
However, on the other hand, a smaller $K$ will also make it easier to infer the input token, thus lowering the privacy protection levels (e.g., for $K = 5$, it does not exceed $0.96$ even when $\epsilon$ has been decreased to $0.001$).

\section{Concluding Remarks}

In this work, we study the problem of differentially private text sanitization.
We propose a novel CusText mechanism consisting of a mapping scheme to assign each input token a customized output set and sampling function generation methods based on the mapping scheme and exponential mechanism to reduce privacy costs while improving the utilities of sanitized texts.
Extensive experiments demonstrate that CusText achieves better privacy-utility trade-offs than state-of-the-art text sanitization mechanisms.
In the future, we will explore how to improve our mechanism by adaptively allocating privacy costs across tokens and find a better way to decide whether a token is sensitive than based on a pre-defined stopword list.

\section*{Acknowledgements}
This work was supported by the National Natural Science Foundation of China (under Grant numbers 62202170, 62202169) and Alibaba Group through the Alibaba Innovation Research Program.

\section*{Limitations}
First, as indicated in \Cref{tb:query}, different tokens are not equally vulnerable to privacy attacks.
As such, assigning every token with the same output size $K$ and privacy parameter $\epsilon$ might not be an ideal choice.
An improved method would be to adaptively allocate privacy costs across tokens so that all of them are adequately protected.
Second, we adopt two simple strategies to decide whether a token is sensitive: assuming all tokens are sensitive or based on a pre-defined stopword list.
However, the prior might be over-protective, but the latter can lead to privacy leakage since stopwords might help infer other sanitized tokens.
Therefore, a more flexible and practical way to decide the sensitivity of tokens is required.

\bibliographystyle{acl_natbib}
\bibliography{custom}

\end{document}